\documentclass[a4paper,11pt]{article}
\pdfoutput=1 
\usepackage{jheppub} 
\usepackage{bbm,bm,graphicx,mathtools,color,hyperref,slashed}
\usepackage{amssymb,amsmath}
\usepackage[T1]{fontenc}

\usepackage[all]{xy}

\graphicspath{{./Figures/}}

\newcommand{\diff}{\mathrm{d}}
\newcommand{\p}{\partial}

\newcommand{\Diff}{{\mathcal{D}}}

\newcommand{\be}{\begin{equation}}      
\newcommand{\ee}{\end{equation}}      
\newcommand{\bea}{\begin{eqnarray}}      
\newcommand{\eea}{\end{eqnarray}}

\newcommand{\tr}{\mathrm{tr}}

\newcommand{\im}{\mathrm{i}}

\newcommand{\calP}{\mathcal{P}}

\newcommand{\rme}{\mathrm{e}}

\newcommand{\ttimes}{\,\tilde{\times}\,}

\title{Modified instanton sum in QCD and  higher-groups}

\author{Yuya Tanizaki, }

\author{Mithat \"Unsal}

\affiliation{North Carlina State University, Raleigh, NC, 27607, USA}

\emailAdd{ytaniza@ncsu.edu}
\emailAdd{unsal.mithat@gmail.com}

\abstract{
We consider the $SU(N)$ Yang-Mills theory, whose topological sectors are restricted to the instanton number with integer multiples of $p$. 
We can formulate such a quantum field theory maintaining locality and unitarity, and the model contains both $2\pi$-periodic scalar and $3$-form gauge fields. This can be interpreted as  coupling a topological theory to Yang-Mills theory, so the local dynamics becomes identical with that of pure Yang-Mills theory.  
The theory has not only $\mathbb{Z}_N$ $1$-form symmetry but also $\mathbb{Z}_p$ $3$-form symmetry, and we study the global nature of this theory from the recent 't~Hooft anomaly matching. 
The computation of 't~Hooft anomaly incorporates an intriguing higher-group structure.
We also carefully examine that how such kinematical constraint is realized in the dynamics by using the large-$N$ and also the reliable semiclassics on $\mathbb{R}^3\times S^1$, and we find that the topological susceptibility plays a role of the order parameter for the $\mathbb{Z}_p$ $3$-form symmetry. 
Introducing a fermion in the fundamental or adjoint representation, we find that the chiral symmetry becomes larger than the usual case by $\mathbb{Z}_p$, and it leads to the extra $p$ vacua by discrete chiral symmetry breaking. 
No dynamical domain wall can interpolate those extra vacua since such objects must be charged under the $3$-form symmetry in order to match the 't~Hooft anomaly.  
}

\begin{document}
\maketitle
\section{Introduction}\label{sec:introduction}

In the path-integral formulation, we should integrate over \textit{all possible} field configurations in order to define a quantum field theory (QFT). 
In many interesting QFTs, the space of field configurations is disconnected and those disconnected components often have different topological numbers. 
It is a fundamental and important question to consider how we sum up different topological sectors when we define a QFT. 

It has been widely believed that the summation over topological sectors is uniquely fixed by unitarity and locality of QFTs~\cite{Belavin:1975fg, Callan:1976je, Jackiw:1976pf, tHooft:1976rip}.  
Especially, the locality, or the cluster-decomposition property, is supposed to require that we must sum up all possible topological sectors. 
For example, in the pure $SU(N)$ Yang-Mills (YM) theory, the gauge field configurations are distinguished by the instanton number, which takes integer values. 
Assume that we are trying to consider the path integral only on the zero-instanton sector. Even in the trivial sector, however, there are configurations with an instanton ($\mathcal{I}$)--anti-instanton $(\overline{\mathcal{I}})$ event. When $\mathcal{I}$ and $\overline{\mathcal{I}}$ are sufficiently separated with each other, the cluster decomposition shows that this produces the same effect of the single instanton event. 
Following the above standard lore, we are summing up all instanton sectors, and multiply the $U(1)$ phase $(\rme^{\im \theta})^\nu$ for $\nu$-instanton sectors, and we obtain the $\theta$ vacua of the Yang-Mills theory.  
This argument clearly shows that the local physics, such as the particle spectra, does not care about the global condition, and this insight is indeed quite useful in order to understand the numerical lattice simulation of gauge theories at the fixed topology~\cite{Brower:2003yx, Aoki:2007ka}.  

However, this does not necessarily mean that QFT without summing up all instantons is ill-defined as a local and unitary QFT~\cite{Pantev:2005zs,Pantev:2005rh, Pantev:2005wj}. 
The current understandings of QFT requires us to specify more data, such as the content of extended operators like the Wilson loop~\cite{Seiberg:2010qd,Hellerman:2010fv, Aharony:2013hda}, in order to uniquely fix the possible summation over topological sectors.  
This affects significantly on the global nature of the theory, especially for the vacuum structure. 

In this paper, we constrain the possible instanton  in various QFTs, including the $SU(N)$  YM theory, ${\cal N}=1$ super YM (SYM) and  quantum chromodynamics (QCD) without changing the local degrees of freedom.  
As shown in Ref.~\cite{Seiberg:2010qd}, this is possible if we sum up the topological sectors whose instanton numbers are divisible by $p\in \mathbb{Z}$. 
In order for this constraint to be consistent with locality, we have to introduce the $2\pi$-periodic scalar field $\chi$ and the $U(1)$ $3$-form gauge field $c^{(3)}$. 
Interestingly, these extra fields do not propagate locally, so the local dynamics is identical with that of the pure Yang-Mills theory. 
However, they can provide the globally nontrivial topological degrees of freedom, and it turns out that we necessarily have the $\mathbb{Z}_p$ $3$-form symmetry, although the local dynamics is  exactly same with that of unmodified 4d QFT.  
Therefore, restricting the instanton sum to multiples of topological charge $p$  
can be interpreted as coupling a  topological quantum field theory (TQFT) to QFT, in particular to  YM and QCD. In this paper, we call them as generalized YM and QCD. 

In the following, let us summarize the main outcomes obtained in this paper.

To understand the physical consequences of modified topological sum, we consider its effect on YM and SYM theories from the viewpoint of 't~Hooft anomaly~\cite{tHooft:1979rat, Frishman:1980dq} (see also Refs.~\cite{Wen:2013oza, Kapustin:2014lwa, Cho:2014jfa, Wang:2014pma, Witten:2016cio, Tachikawa:2016cha, Gaiotto:2017yup, Tanizaki:2017bam, Kikuchi:2017pcp, Komargodski:2017dmc, Komargodski:2017smk, Shimizu:2017asf, Wang:2017loc,Gaiotto:2017tne, Tanizaki:2017qhf, Tanizaki:2017mtm,  Yamazaki:2017dra, Guo:2017xex,  Sulejmanpasic:2018upi, Tanizaki:2018xto, Yao:2018kel, Kobayashi:2018yuk,  Tanizaki:2018wtg,Anber:2018jdf,   Anber:2018xek, Armoni:2018bga, Yonekura:2019vyz, Nishimura:2019umw, Misumi:2019dwq, Cherman:2019hbq} for recent advances) and semiclassics on $\mathbb{R}^3\times S^1$~\cite{Unsal:2007jx,Unsal:2007vu,Unsal:2008ch, Poppitz:2011wy, Poppitz:2012sw,Poppitz:2012nz, Argyres:2012vv,Argyres:2012ka, Anber:2011gn, Anber:2015wha}. 
Because of the modified sum over instantons, the periodicity of Yang-Mills topological angle, $\theta_{\mathrm{YM}}$, is shortened as $\theta_{\mathrm{YM}}\sim \theta_{\mathrm{YM}}+{2\pi\over p}$ instead of the usual $2\pi$ periodicity. 
In Sec.~\ref{sec:vacuum_mYM}, we find that the $1$-form symmetry $\mathbb{Z}_{N}^{[1]}$ and the $3$-form symmetry $\mathbb{Z}_{p}^{[3]}$ forms an intriguing $4$-group structure, and we denote it as $\mathbb{Z}_{N}^{[1]}\ttimes \mathbb{Z}_{p}^{[3]}$ to emphasize that it is not a direct product. 
By introducing the appropriate background gauge field for $\mathbb{Z}_{N}^{[1]}\ttimes \mathbb{Z}_{p}^{[3]}$, we find that this periodicity is extended to $2\pi N$ periodicity. As a consequence, we can expect that the generalized YM theory has $Np$ branch structure. 
This expectation is confirmed by the large-$N$ counting and by the semiclassical computations.  We also find that the $\mathbb{Z}_{p}^{[3]}$ symmetry is unbroken so long as the topological susceptibility $\mathcal{X}_{\mathrm{top}}$ of the YM theory is nonzero.

We also discuss  compatibility of  abelian duality and semi-classics with  mixed anomaly.   In this context, 
we consider center-stabilized YM theory on small $\mathbb R^3 \times S^1$, a theory that admits a reliable  semi-classical treatment, in which confinement and mass gap can be shown  \cite{Unsal:2008ch}  by using similar techniques with Polyakov model \cite{Polyakov:1975rs}. 
 In particular, we prove that the abelian dual formulation based on dilute gas of monopole-instantons of YM 
 produces the same mixed anomaly polynomial \cite{Gaiotto:2017yup}  as  the original Yang-Mills theory, and $Np$ branched vacuum structure.    This provides further evidence for the adiabatic continuity idea \cite{Unsal:2008ch}.

As we will show in Sec.~\ref{sec:vaccum_mSYM}, for SYM with modified instanton sum,  the discrete chiral symmetry becomes $\mathbb{Z}_{2Np}$, so it is greater than the usual case by $\mathbb{Z}_p$. 
As a correspondence to the existence of $Np$-branch structure in pure generalized YM theory, the discrete chiral symmetry is spontaneously broken to $\mathbb{Z}_2$, so the theory has $Np$ vacua.   There is a subtle issue related to these $Np$-vacua, since our QFT has a TQFT component, which we explain. The $N$ vacua related by the usual discrete chiral transformation, which corresponds to $\theta_{\mathrm{YM}}\to \theta_{\mathrm{YM}}+2\pi$, show the same dynamical behaviors as in the usual SYM. 
On the other hand, other vacua, related by $\theta_{\mathrm{YM}}\to \theta_{\mathrm{YM}}+{2\pi\over p}$, turn out to have a distinct feature: There is no domain wall as a dynamical excitation interpolating between them!  
A mixed anomaly between chiral and three-form symmetry shows that such wall-type excitation must be charged under the three-form symmetry. Thus, such configuration is not allowed unless we insert the $3$-volume operator, the order parameter for the three-form symmetry, in the path integral. 
This leads to much stronger superselection rule than what we usually think of in the local QFT, and we propose to call such distinct vacua as different universes\footnote{ In a related discussion in a 2d QFT,  Zohar Komargodski  emphasized 
 the distinction between various types of vacua.  He referred to vacua which obey much stronger superselection rule as 
universes. The authors thank Zohar Komargodski for proposing this nice name and helpful discussions.}. 
 In other words, breaking of the $\mathbb{Z}_{p}\simeq \mathbb{Z}_{2Np}/\mathbb{Z}_{2N}$ part of chiral symmetry  is due to the mixed anomaly with $\mathbb{Z}_p^{[3]}$, leading to distinct universes. It has the different physical origin from the breaking of $\mathbb{Z}_{2N}$, since the spontaneous breaking $\mathbb{Z}_{2N}\to \mathbb{Z}_2$ is a dynamical IR phenomenon as in usual SYM, leading to superselection sectors within each universe.
The difference between the usual superselection sectors and universes is discussed in details in Sec.~\ref{sec:superselection_universes}.

We also consider, in Sec.~\ref{sec:QCD_fund}, the modified instanton sum in QCD with one-flavor fundamental quark. 
In the usual case, the massless point does not have any axial symmetry because of quantum anomaly~\cite{Adler:1969gk, Bell:1969ts}. 
In the case of modified sum, we obtain the discrete chiral symmetry $\mathbb{Z}_{2p}$ as a combination of two quantum-mechanically broken $U(1)$ symmetries, $U(1)$ axial symmetry and Peccei-Quinn $U(1)$ symmetry. 
We show the mixed 't~Hooft anomaly for chiral symmetry, $\mathbb{Z}_{2p}$, and three-form symmetry, $\mathbb{Z}_{p}^{[3]}$, and find that there are $p$ distinct universes.

In this paper, we focus on the effect of modified instanton sum in $4$d gauge theories. In the case of $2$d field theories,  
these constructions are equivalent to charge-$p$ sigma models on gerbes  \cite{Pantev:2005zs,Pantev:2005rh, Pantev:2005wj}.  Our construction provides an interpretation for the enriched vacuum structure obtained in these works,  which  
exhibit a  very similar structures of vacua to $4$d  generalized gauge theory.   In order to uncover the reason for this, in Sec.~\ref{sec:4d_2d}, we prove that these theories share the same anomaly structure so that it is quite natural to have the same vacuum structures.

\section{Modification of instanton sum in Yang-Mills theory}\label{sec:modification}

In this section, we first give a brief review on the $SU(N)$ Yang-Mills theory, and consider the restriction of the instanton number modulo $p$ following \cite{Seiberg:2010qd}.  This will give a generalized version of Yang-Mills theory which is locally the same  but globally different compared to original theory.

\subsection{Instanton sum in Yang-Mills theory}

Let $a$ be the dynamical $SU(N)$ gauge field, and the classical action of the Yang-Mills theory is given by 
\be
S_{\mathrm{YM}}={1\over 2g^2_{\mathrm{YM}}}\int \tr[F(a)\wedge \star F(a)]+{\im\, \theta_{\mathrm{YM}}\over 8\pi^2}\int \tr[F(a)\wedge F(a)]. 
\ee
In our convention, $a$ is locally a Hermitian traceless matrix ($\mathfrak{su}(N)$)-valued one-form, and the field strength is given by 
\be
F(a)=\diff a+\im\, a\wedge a. 
\ee
For simplicity, the spacetime manifold is restricted to closed four-dimensional spin manifolds throughout this paper. 

On closed four-dimensional manifold, the instanton number is quantized as 
\be
{1\over 8\pi^2}\int \tr[F\wedge F]=\nu\in \mathbb{Z}. 
\ee
This topological number is determined completely by the transition function of the principal $SU(N)$ bundle \cite{vanBaal:1982ag}. 
When the instanton numbers of two gauge field configurations are the same, we can perform the gauge transformation so that the transition functions of those field configurations are the identical. 
Therefore, the path integral on the fixed instanton sector, $Z_\nu$,  can be obtained as the integration over all possible $\mathfrak{su}(N)$-valued one-form field $\delta a$, 
\be
Z_\nu= \int \Diff[\delta a] \exp \left(-{1\over 2g^2_{\mathrm{YM}}} \int \tr[F(a_\nu+\delta a)\wedge \star F(a_\nu+\delta a)]\right), 
\ee
where $a_\nu$ is an arbitrary reference gauge field with the instanton number $\nu$. 
We define the theta vacua by summing up all the instanton sectors as 
\be
Z(\theta_{\mathrm{YM}})=\sum_{\nu \in \mathbb{Z}} \rme^{-\im\, \nu\, \theta_{\mathrm{YM}}} Z_\nu. 
\label{eq:usual_instanton_sum}
\ee
By definition, there is the $2\pi$ periodicity of the theta angle, $\theta_{\mathrm{YM}}\sim \theta_{\mathrm{YM}}+2\pi$. 

Why do we have to sum up all the instanton sectors? 
The answer comes out of the locality and the unitarity of QFT~\cite{Belavin:1975fg, Callan:1976je, Jackiw:1976pf, tHooft:1976rip} (see also \cite{Weinberg2}). 
The locality of QFT requires that if the closed spacetime is given by the disjoint union then the partition function is 
\be
Z_{M_1\sqcup M_2}= Z_{M_1} Z_{M_2}. 
\ee
The unitarity is translated as the reflection positivity, so when we flip the orientation of the spacetime, denoted as $-M$, the partition function is complex conjugated, 
\be
Z_{-M}= \overline{Z_M}. 
\ee
Because of the fact that there is a configuration with any given instanton number on any closed $4$-manifolds, we get
\be
Z_{\nu, \, M_1\sqcup M_2}=\sum_{\nu' \in \mathbb{Z}} Z_{\nu-\nu',\, M_1} Z_{\nu', \,M_2}, 
\ee
where $Z_{\nu, M}$ is the path integral over $M$ at the $\nu$-instanton sector. Therefore, (\ref{eq:usual_instanton_sum}) is the unique form of solutions that is consistent with locality and unitarity, when $Z_{\nu, M}\not =0$ for any $\nu\in \mathbb{Z}$. 

\subsection{Modifying the summation over topological sectors}

Now, instead of performing the path integral over all possible instanton sectors as in (\ref{eq:usual_instanton_sum}), let's consider the modified instanton sum.  
It was a lore that such a procedure breaks the cluster decomposition property of QFT, because the instanton number is globally defined. 
However, recently, it is uncovered that this is too strict as a requirement of local QFT~\cite{Seiberg:2010qd}.

The trick is to introduce the dynamical $U(1)$ three-form gauge field $c^{(3)}$ and the $2\pi$-periodic scalar field $\chi$. The action of the theory is given as follows:
\bea
S_{\mathrm{gYM}}&=&{1\over 2g^2}\int \tr[F(a)\wedge \star F(a)]+{\im \theta_{\mathrm{YM}}\over 8\pi^2}\int \tr[F(a)\wedge F(a)]\nonumber\\
&&+\im \int \chi\wedge \left({1\over 8\pi^2}\tr[F(a)\wedge F(a)]-{p\over 2\pi}\diff c^{(3)}\right)+{\im \hat{\theta}\over 2\pi}\int \diff c^{(3)}, 
\label{eq:modified_YM}
\eea
and we call this as generalized YM. There are two alternative ways to think about it as follows. 

The equation of motion of $\chi$ gives 
\be
{1\over 8\pi^2}\tr[F(a)\wedge F(a)]={p\over 2\pi}\diff c^{(3)}. 
\label{eq:constrained_instanton}
\ee
Therefore, the equation of motion can be solved if and only if the instanton number is divisible by $p$. 
Since the constraint equation (\ref{eq:constrained_instanton}) is local, the locality of QFT is manifest. 
Also, the local fluctuation of $c^{(3)}$ is completely constrained by (\ref{eq:constrained_instanton}) up to the gauge transformation, so $c^{(3)}$ does not have the locally propagating degree of freedom. 

 Alternatively,  the equation of motion of $c^{(3)}$ is locally given by 
\be
\diff \chi=0,
\ee
so $\chi$ does not have the propagating degree of freedom either.   To see the restriction to charge-$p$ sector,  
we note that  the possible values of 
$\chi$ are given by $
\chi={2\pi\over p}\ell, \;  \ell=0,1,\ldots,p-1$. The sum over $\ell$  for   sectors with instanton numbers  $k\neq 0$ mod $p$  vanishes  leading to the same conclusion that only sectors with instanton number  divisible by $p$ survive. 

Since neither $\chi$ nor $c^{(3)}$ are propagating degrees of freedom, 
 the local dynamics of generalized YM, (\ref{eq:modified_YM}), is identical with that of the usual Yang-Mills theory.

Using (\ref{eq:constrained_instanton}), we can combine two theta terms into the one term as 
\be
{\im \theta_{\mathrm{YM}}\over 8\pi^2}\int \tr[F(a)\wedge F(a)]+{\im \hat{\theta}\over 2\pi}\int \diff c^{(3)}={\im (\hat{\theta}+p\, \theta_{\mathrm{YM}})\over 2\pi}\int \diff c^{(3)}. 
\ee
Therefore, we can set one of them to be zero without loss of generality. The periodicity of $\hat{\theta}$ is $2\pi$, and we can relate it to the usual Yang-Mills theta angle by 
\be
\hat{\theta} = p \, \theta_{\mathrm{YM}}.
\ee 
In the conventional $\theta_{\mathrm{YM}}$ angle, the periodicity looks to be shortened to $2\pi/p$ from $2\pi$. 
This is the consequence of the fact that we are only summing up the instanton sectors with multiples of $p$. 

\section{Vacuum structures of generalized Yang-Mills theory}\label{sec:vacuum_mYM}

In this section, we study the physical consequences out of the restriction of instanton sum. We first identify the symmetry of the generalized YM, and observe that there is $\mathbb{Z}_{Np}$ mixed anomaly between the higher-form symmetries and the $\theta$-angle periodicity. 
This naturally leads to the existence of $Np$-branch structure, and we explicitly confirm it by the large-$N$ discussion and also by the semiclassical analysis on $\mathbb{R}^3\times S^1$ with double-trace deformation. 

\subsection{Extended operators and higher-form symmetry}

Since we have the extra field contents, the theory is expected to have the larger set of symmetry. Indeed, the theory (\ref{eq:modified_YM}) has the $1$-form symmetry $\mathbb{Z}_{N}^{[1]}$ and $3$-form symmetry $\mathbb{Z}_{p}^{[3]}$, and we denote it as 
\be
\mathbb{Z}_{N}^{[1]} \ttimes \mathbb{Z}_{p}^{[3]}. 
\ee 
We use a special symbol $\ttimes$ instead of $\times$, because the group structure may not be the naive direct product but contains a nontrivial extension. In the next subsection, we will discuss that this possible extension indeed occurs, and find the $4$-group structure appears. 
Here, in order to identify each ingredient of the symmetry, $\mathbb{Z}_{N}^{[1]}$ and $\mathbb{Z}_{p}^{[3]}$, we first list the relevant gauge invariant operators, and check the commutation relation by solving the equation of motion. 

\noindent
\textbf{Point-like operators: } 
In addition to the usual point-like operators, $\tr[F_{\mu\nu}F_{\rho\sigma}]$, in the pure Yang-Mills theory, the generalized YM theory has the $U(1)$-valued scalar field,
\be
U_0(x)=\rme^{\im \chi(x)}. 
\ee
It is notable that the equation of motion of $c^{(3)}$ gives $p\, \diff \chi=0$, and thus this operator is the codimension-$4$ topological defect. We shall see that this is the generator of $\mathbb{Z}_p^{[3]}$. 

\noindent
\textbf{Line operators: } The theory has the Wilson line operators along the closed line $M_1$, 
\be
W(M_1) = \mathrm{tr}\left[\calP \exp \left(\im \int_{M_1}a \right)\right]. 
\ee
This is the charged object under $\mathbb{Z}_{N}^{[1]}$. 

\noindent
\textbf{Surface operators: } In the pure Yang-Mills theory, there is the topological codimension-$2$ defect, $U_2(M_2)$~\cite{Gaiotto:2014kfa, Gukov:2008sn}. This operator can be regarded as the world sheet of the Dirac string emitted from the 't~Hooft loop operator~\cite{tHooft:1977nqb}. 
As a consequence of the Wilson-'t~Hooft commutation relation, 
\be
\Bigl\langle U_2(M_2) W(M_1)\Bigr\rangle = \exp\left({2\pi\im\over N}\mathrm{Link}(M_2,M_1)\right) \Bigl\langle W(M_1)\Bigr\rangle. 
\ee
Therefore, $U_2(M_2)$ is the generator of $\mathbb{Z}_{N}^{[1]}$ symmetry that acts on $W(M_1)$. 

\noindent
\textbf{Volume operator: } We can consider the Wilson volume operator on closed $3$-manifolds $M_3$ as 
\be
V^{(3)}(M_3)= \exp\left(\im \int_{M_3} c^{(3)}\right). 
\ee
This is the charged object under $\mathbb{Z}_{p}^{[3]}$. Indeed, solving the equation of motion of $\chi$, we obtain 
\be
\Bigl\langle \rme^{\im \chi(x)} V^{(3)}(M_3)\Bigr \rangle =\exp\left({2\pi\im\over p} \mathrm{Link}(x, M_3)\right)\Bigl\langle V^{(3)}(M_3)\Bigr \rangle,
\ee
which shows that $\mathbb{Z}_{p}^{[3]}$ is generated by $U_0(x)$. 

\subsection{Mixed anomaly between $\mathbb{Z}_{N}^{[1]}\ttimes \mathbb{Z}_{p}^{[3]}$ symmetry and the $\theta$-angle periodicity, and 4-group structure}
\label{sec:mixed_anomaly_theta}

In order to figure out the vacuum structure of the generalized YM, we study the generalized 't~Hooft anomaly, or the global inconsistency, between $\mathbb{Z}_{N}^{[1]}\ttimes \mathbb{Z}_{p}^{[3]}$ symmetry and the periodicity of $\theta$ angle. 
In the process of gauging, we encounter an interesting $4$-group structure. 

Let us first try to gauge the $\mathbb{Z}_{N}$ $1$-form symmetry. 
For this purpose, we realize the $\mathbb{Z}_{N}$ two-form gauge field as a pair of $U(1)$ one-form and two-form gauge fields, $B^{(1)}$, $B^{(2)}$, with the constraint, 
\be
N B^{(2)} = \diff B^{(1)}. 
\ee
Following Refs.~\cite{Kapustin:2014gua, Gaiotto:2014kfa}, we introduce the $U(N)$ gauge field, $\widetilde{a}$, and relate it to the dynamical $SU(N)$ gauge field $a$ locally as 
\be
\widetilde{a}=a+{1\over N}B^{(1)}. 
\ee
Since each term on the right hand side does not have a gauge-invariant meaning, this expression should be understood as a formal expression, and it means that the path integral $\int \Diff \widetilde{a}$ is performed over the $U(N)$ gauge fields with the constraint 
\be
\tr\left[\widetilde{a}\right]=B^{(1)}.
\ee 
In order to gauge $\mathbb{Z}_{N}^{[1]}$ correctly, we have to postulate the invariance under the one-form gauge transformation, 
\be
B^{(2)}\mapsto B^{(2)}+\diff \Lambda^{(1)},\; B^{(1)}\mapsto B^{(1)}+N\Lambda^{(1)},  
\ee
and 
\be
\widetilde{a}\mapsto \widetilde{a}+\Lambda^{(1)}.
\ee
In the case of pure Yang-Mills theory, we can do this by replacing the $SU(N)$ field strength $F(a)$ in the Lagrangian by the gauge-invariant combination of the $U(N)$ field strength $\widetilde{F}=F(\widetilde{a})$ and $B^{(2)}$: 
\be
F(a)\Rightarrow \widetilde{F}-B^{(2)}. 
\ee
This operation produce  the mixed anomaly between $\theta$-angle periodicity and $\mathbb Z_N$ 1-form symmetry, i.e, 
\be
Z_{\rm YM} (\theta_{\rm YM} + 2\pi)=  {\rm exp} \left( {\rm i}  { N\over 4 \pi }\int B^{(2)}\wedge B^{(2)} \right)Z_{\rm YM} (\theta_{\rm YM})
\ee
because of the quantization condition:
\be
 {N\over 8\pi^2}\int B^{(2)}\wedge B^{(2)}\in  \frac {1}{N} \mathbb{Z}.
 \label{q1}
\ee
corresponding to a non-trivial 't Hooft flux.

However,  this operation  does not work  in the generalized YM theory.  Equation of motion of  Lagrange multiplier $\chi$ gives 
\be
{1\over 8\pi^2}(\tr\widetilde{F}^2-N (B^{(2)})^2)+{p\over 2\pi}\diff c^{(3)}=0. 
\ee
This automatically requires that 
\be
{N\over 8\pi^2}\int B^{(2)}\wedge B^{(2)}\in \mathbb{Z}, 
\ee
in sharp contrast with \eqref{q1}.  Hence, the nontrivial 't~Hooft magnetic flux cannot be consistently introduced as the background field.  This shows that we cannot gauge the $\mathbb{Z}_{N}^{[1]}$ by itself.

We can resolve this puzzle by gauging $\mathbb{Z}_{p}^{[3]}$ at the same time. In order to gauge it, we introduce the $U(1)$ $3$-form and $4$-form gauge fields $D^{(3)}$ and $D^{(4)}$. The $3$-form gauge transformation is defined by 
\be 
D^{(4)}\mapsto D^{(4)}+\diff \Lambda^{(3)},\; D^{(3)}\mapsto D^{(3)}+p\Lambda^{(3)}, 
\ee
and 
\be
c^{(3)}\mapsto c^{(3)}+\Lambda^{(3)}. 
\ee
The minimal coupling procedure for the Lagrange-multiplier term gives
\be
{\im\over 2\pi}\chi\wedge \left[{1\over 4\pi}\tr \widetilde F^2-p\, \diff c^{(3)}+\diff D^{(3)}\right]. 
\ee
This is clearly $U(1)$ three-form gauge invariant.
At the first sight, however, this may not seem to be invariant under the 1-form gauge transformation because 
$\widetilde F$ transforms non-trivially. 
We can establish it by requiring that $D^{(3)}$ transforms under the one-form gauge transformation as 
\be
D^{(3)}\mapsto D^{(3)}-\left({N\over 2\pi}B^{(2)}\wedge \Lambda^{(1)}+{N\over 4\pi}\Lambda^{(1)}\wedge \diff \Lambda^{(1)}\right). 
\ee
Since the right hand side has the correct periodicity as the $U(1)$ three-form gauge field,   i.e., 
$\int_{M_4} \diff ( \ldots) \in 2 \pi \mathbb Z$,   this is a well-defined transformation. 
Notice that the equation of motion of $\chi$ does no longer  give any constraint on $B^{(2)}$, and thus the nontrivial magnetic flux is now allowed to be inserted.  
Because of this extra one-form transformation, the pair $(D^{(3)}, D^{(4)})$ is no longer the naive $\mathbb{Z}_{p}$ four-form gauge field, but it is affected by the $\mathbb{Z}_{N}$ two-form background gauge field. 
The constraint equation turns out to be 
\be
p D^{(4)}=\diff D^{(3)}+{N\over 4\pi}B^{(2)}\wedge B^{(2)}. \label{eq:4group}
\ee
As a consequence, 
\be
 \int D^{(4)}={2\pi\over Np}\Biggl(\underbrace{N\int  {\diff D^{(3)}\over 2\pi}}_{\in  N\mathbb{Z}}+\underbrace{{N^2\over 8\pi^2}\int (B^{(2)})^2}_{\in \mathbb{Z}}\Biggr)\in {2\pi\over Np}\mathbb{Z}. 
\ee
This expression shows that at the intersection of two generic codimension-$2$ defects $U_2$, we must assign the $\mathbb{Z}_{Np}$ phase, and the one-form symmetry is transmuted to the three-form symmetry. 
The analogous situation also appears between the $0$-form and $1$-form symmetry, which leads to the $2$-group structure~\cite{Sharpe:2015mja, Cordova:2018cvg, Benini:2018reh}. 
Accordingly, ours is an example of the $4$-group structure in $4$-dimensional 
QFT\footnote{
This turns out to be a special case of the general theorem discussed in Sec.~3.2 of Ref.~\cite{Tachikawa:2017gyf}. 
We consider a $d$-dim QFT that has the global symmetry $\mathbb{Z}_{N}^{[n]}\times \mathbb{Z}_{p}^{[m]}$ with mixed 't~Hooft anomaly. By gauging the non-anomalous subgroup $\mathbb{Z}_{p}^{[m]}$, the resultant theory has the remnant symmetry $\mathbb{Z}_N^{[n]}$ and acquires the dual symmetry $\mathbb{Z}_{p}^{[d-m-2]}$. 
As a consequence of the mixed 't~Hooft anomaly in the original theory, the group structure between $\mathbb{Z}_N^{[n]}$ and $\mathbb{Z}_p^{[d-m-2]}$ needs a nontrivial extension, $\mathbb{Z}_N^{[n]}\ttimes \mathbb{Z}_{p}^{[d-m-2]}$, which forms a $(d-1-m)$-group structure. 
In our case, $\mathbb{Z}_{N}^{[n]}=\mathbb{Z}_{N}^{[1]}$ is the center symmetry of $SU(N)$ pure Yang-Mills theory, and $\mathbb{Z}_{p}^{[-1]}\subset U(1)^{[-1]}$ is a subgroup of ``(-1)-form symmetry'' that means the $2\pi$-periodicity of the $\theta$ parameter. 
This argument clarifies that the $4$-group structure in this paper can be interpreted as a consequence of the generalized mixed anomaly, or global inconsistency, in pure Yang-Mills theory. 

The authors thank the anonymous referee pointing out the above arguments, clarifying the underlying structure. 
}.

The gauged action is 
\bea
S_{\mathrm{gauged}}&=&{1\over 2g^2}\int \tr|\widetilde{F}-B^{(2)}|^2+{\im \theta_{\mathrm{YM}}\over 8\pi^2}\int \tr[(\widetilde{F}-B^{(2)})^2]\nonumber\\
&&+\im \int \chi\wedge \left({1\over 8\pi^2}\tr[\widetilde{F}^2]-{p\over 2\pi}\diff c^{(3)}+\diff D^{(3)}\right)+{\im \hat{\theta}\over 2\pi}\int (\diff c^{(3)}-D^{(4)}).  
\eea
Thus, the shift $\theta_{\mathrm{YM}}\to \theta_{\mathrm{YM}}+2\pi/p$ ($\Leftrightarrow \hat{\theta}\to \hat{\theta}+2\pi$) gives the extra phase of the partition function by 
\be
Z(\theta_{\mathrm{YM}}+2\pi/p)=\exp\left(-\im \int D^{(4)}\right) Z(\theta_{\mathrm{YM}}), 
\ee
and the overall $\mathbb{Z}_{Np}$ phase is characterized by the background gauge fields. 
When $\theta_{\mathrm{YM}}$ is  quantized in an odd-integer multiple  of $\pi/p$ (half of its domain $2\pi/p$), this relation can be used to obtain the mixed 't~Hooft anomaly or the global inconsistency between $CP$ symmetry and $\mathbb{Z}_{N}^{[1]}\ttimes \mathbb{Z}_p^{[3]}$, generalizing ~\cite{Gaiotto:2017yup, Tanizaki:2017bam, Kikuchi:2017pcp, Komargodski:2017dmc,  Karasik:2019bxn} to cases  involving 4-group symmetry.  
Recently, this type of relation (for 1-form symmetry)   itself is understood as the generalized 't~Hooft anomaly with the $\theta$-angle periodicity~\cite{Cordova:2019jnf}, and it naturally leads to the existence of multi-branch structure as will be discussed in the following subsections. 

\subsection{Large-$N$ limit}

In order to get an insight how the anomaly is matched by low-energy physics, we study its dynamics by using the large-$N$ discussion in this subsection and the semiclassical analysis on $\mathbb{R}^3\times S^1$ in the next subsection. 

Here, we set $\hat{\theta}=0$ and discuss the physics of $\theta_{\mathrm{YM}}$. 
We first perform the path integral over the $SU(N)$ gauge field following the standard large-$N$ counting~\cite{Witten:1980sp, Witten:1998uka}. Introducing the 't~Hooft coupling $\lambda_t=g^2 N$, the Yang-Mills action can be written as 
\be
S_{\mathrm{YM}}=N\int \left[{1\over 2\lambda_t}\tr |F|^2+\im {\theta_{\mathrm{YM}}+\chi\over N}\wedge  {1\over 8\pi^2}\tr(F^2)\right]. 
\ee
Therefore, the natural angle parameter in the large-$N$ limit is given by $\theta_{\mathrm{YM}}/N$ instead of $\theta_{\mathrm{YM}}$ itself, and the energy density of a ground state is given by 
\be
\mathcal{E}=N^2 f\left({\theta_{\mathrm{YM}}+\chi\over N}\right)
\simeq {\mathcal{X}_{\mathrm{top}}\over 2} (\theta_{\mathrm{YM}}+\chi)^2, 
\ee
where $\mathcal{X}_{\mathrm{top}}$ is the topological susceptibility. The overall factor $N^2$ counts the number of local degrees of freedom, and $f$ is an $N$-independent function. 
Since this prescription breaks the $2\pi$ periodicity of the pure YM theory, we have to recover it by introducing the multi-branch structure~\cite{Witten:1980sp, Witten:1998uka}. They are labeled by $k\in \mathbb{Z}$ with the energy density 
\be
\mathcal{E}_k={\mathcal{X}_{\mathrm{top}}\over 2} (\theta_{\mathrm{YM}}+\chi+2\pi k)^2. 
\ee
The partition function of the Yang-Mills theory is then given by 
\be
Z_{\mathrm{YM}}(\theta_{\mathrm{YM}},\chi)=\sum_{k} \exp(- \mathrm{Vol} \,\mathcal{E}_k(\theta_{\mathrm{YM}},\chi)), 
\ee
where $\mathrm{Vol}$ is the four volume of the spacetime. 

In order to obtain the partition function of the generalized YM, we next perform the path integral over $\chi$ and $c^{(3)}$:
\be
Z_{\mathrm{gYM}}=\int \Diff\chi \Diff c^{(3)} \exp\left(-\im{p\over 2\pi}\int \chi\wedge \diff c^{(3)}\right) Z_{\mathrm{YM}}(\theta_{\mathrm{YM}},\chi). 
\ee
The path integral over $c^{(3)}$ can be performed exactly, and it gives a delta functional constraint on $\chi$ so that 
\be
\chi={2\pi\over p}\ell,\qquad (\ell=0,1,\ldots,p-1). 
\ee
As a result, we obtain 
\be
Z_{\mathrm{gYM}}(\theta_{\mathrm{YM}})=\sum_{k\in \mathbb{Z}} \sum_{\ell=0}^{p-1}\exp\left(- \mathrm{Vol} {\mathcal{X}_{\mathrm{top}}\over 2} \left(\theta_{\mathrm{YM}}+{2\pi \over p}\ell +2\pi k\right)^2\right),
\ee
and we correctly find the $2\pi/p$-periodicity in terms of $\theta_{\mathrm{YM}}$. 
In order to characterize the multi-branch, we have introduced the two integer, $\ell\sim \ell+p$ and $k\sim k+N$, while the periodicity of $k$ cannot be seen in the large-$N$ limit. 
In the expression of the (quasi) ground-state energies, the periodicity of these labels should have the following extension,
\be
(\ell+p, k)\sim (\ell, k+1). 
\ee
This relation is nothing but the realization of the relation between the four-form and two-form background gauge fields given in (\ref{eq:4group}). 

We can also show that $\mathbb{Z}_{p}$ $3$-form symmetry is unbroken for generic $\theta_{\mathrm{YM}}$. For simplicity, let us consider the case $\theta_{\mathrm{YM}}=0$. 
In order to judge if the three-form symmetry is broken or not, we compute the expectation value, 
\be
\Bigl \langle V^{(3)}(M_3) \Bigr \rangle = \Bigl \langle \exp\left(\im \int_{M_3}c^{(3)}\right) \Bigr\rangle,   
\ee
in the limit $M_3$ becomes larger. If $\langle V^{(3)}\rangle \to 0$ in that limit after appropriate renormalization, then the three-form symmetry is unbroken. 
Let us rewrite this operator as 
\be
V^{(3)}(M_3) = \exp\left(\im \int_{V_4} \diff c^{(3)} \right), 
\ee
where $V_4$ is a subregion of the four-dimensional spacetime bounded by  $M_3$, $\p V_4=M_3$. Therefore, insertion of $V^{(3)} (M_3)$ shifts the scalar field $\chi$ by $2\pi/p$ if it is inside of $M_3$, while fields outside of $M_3$ are not affected. 
Since we take the vacuum configuration outside of $M_3$, the energy density inside of $M_3$ is given by 
\be
\mathcal{E}(2\pi/p)-\mathcal{E}(0)={\mathcal{X}_{\mathrm{top}}\over 2}\left({2\pi\over p}\right)^2, 
\ee
and thus 
\be
\Bigl \langle V^{(3)}(M_3) \Bigr \rangle \simeq \exp\left(-{2\pi^2\mathcal{X}_{\mathrm{top}}\over p^2}\, \mathrm{Vol}(V_4)\right). 
\ee
This shows that, in the large-$N$ limit, the $\mathbb{Z}_p$ three-form symmetry is a symmetry with an order parameter $\mathcal{X}_{\mathrm{top}}$. 
Therefore, in the large-$N$, we obtain the following criterion, 
\begin{align}
&\mathcal{X}_{\mathrm{top}}=  0\quad  \Leftrightarrow \quad \mathbb{Z}_p^{[3]} \xrightarrow{\mathrm{SSB}} 1,  \cr
&\mathcal{X}_{\mathrm{top}} \neq  0\quad  \Leftrightarrow \quad \mathbb{Z}_p^{[3]} \;\;  {\rm unbroken} .
\label{eq:order_3form}
\end{align}
In the YM, $\mathcal{X}_{\mathrm{top}}\sim \Lambda^4$, where $\Lambda$ is the strong scale, so the three-form symmetry of generalized YM is unbroken.   We expect that in $d$-dimensional QFT,   $(d-1)$-form symmetries cannot generally  be broken unless a mixed anomaly requires it, and this is consistent with that expectation.
The above discussion assumes that the large-$N$ counting works nicely, and we note that the numerical lattice simulation supports its correctness (see, e.g., Ref.~\cite{Bonati:2016tvi}). 
In the later sections, we shall see that the three-form symmetry is spontaneously broken if the massless fermion is introduced, and this is consistent with the above observation. 

\subsection{Semiclassics with double-trace deformation on $\mathbb{R}^3\times S^1$ vs. mixed anomaly}

In the previous subsection, we discuss the dynamics of generalized YM in the large-$N$ limit. 
The large-$N$ counting provides a useful picture about the $\theta$-dependence, but the discussion there is qualitative because the theory is a matrix-like model. 
In order to get more quantitative understanding, we consider the small $S^1$ compactification with size $L$, and we add the double trace term in order to force the confinement of Polyakov loops~\cite{Unsal:2007jx,Unsal:2007vu,Unsal:2008ch}. 
The following semiclassical analysis turns out to be reliable without any infrared divergences if the condition $NL\Lambda\ll 1$ is satisfied. 
Below, we set $\hat{\theta}=0$ and simply denote $\theta_{\mathrm{YM}}=\theta$. 
The notation of the $\mathfrak{su}(N)$ Lie algebra used below is summarized in Appendix~\ref{sec:su(N)_root_weight}. 

By $S^1$ compactification, the Wilson loop wrapping around $S^1$ provides the compact adjoint scalar field, $W(S^1)=\exp(\im \bm{\phi})$, and we take the Polyakov gauge so that $\bm{\phi}$ is in the Cartan subalgebra. In components, we denote it as $\bm{\phi}=\sum_{i=1}^{N-1}\phi_i \bm{\alpha}_i=(\phi_1,\phi_2-\phi_1,\ldots, -\phi_{N-1})$ with $2\pi$-periodic fields $\phi_i$. 
At generic values of $\bm{\phi}$, the off-diagonal $3$d gluon fields $a_{ij}$ gets the mass term $|\bm{\phi}\cdot \bm{\alpha}_{ij}|/L$, and the adjoint Higgsing occurs so that the gauge group is broken to $SU(N)\to U(1)^{N-1}$ (and the Weyl group $W_{\mathfrak{su}(N)}$). 

In the case of pure YM, the one-loop effective potential of $W(S^1)$ favors the center-broken minima~\cite{Gross:1980br}, $\bm{\phi}=0$, and the adjoint Higgsing does not occur. 
In order to study the physics of the confined phase, we must add certain deformations, so that the minima is given by the center-symmetric one, $\rme^{\im \bm{\phi}_\star}=(1,\omega,\omega^2,\ldots, \omega^{N-1})$ up to $W_{\mathfrak{su}(N)}$, with $\omega=\rme^{2\pi\im/N}$. 
For this purpose, we add the double-trace deformation to the effective potential, $ \sum_{k=1}^{ \lfloor  {N \over 2}  \rfloor} 
|\tr\, W(S^1)^k|^2$. This  deformation arise naturally from a microscopic theory, QCD with multi-flavor adjoint fermion with  mass  term endowed with  the periodic boundary conditions, and  it may be considered as a low energy effective description. 
 Then, all the off-diagonal gluons gets the mass $\ge {2\pi/NL}$, and below its energy scale the perturbative physics is completely Abelianized, which is sometimes called the classical Coulomb branch. We denote the diagonal gluons as $\bm{a}=\sum_i a_i \bm{\alpha}_i$ with canonically normalized $U(1)$ gauge fields $a_i$. 

We first analyze the YM sector. The effective Yang-Mills action on $\mathbb{R}^3\times S^1$ at the classical Coulomb branch is given by 
\be
S_{\mathrm{YM}}=
 \int_{\mathbb{R}^3} \left({1\over 2g^2 L}|\diff \bm{\phi}|^2+{L\over 2g^2}|\diff \bm{a}|^2+{\im \theta\over 4\pi^2}\diff \bm{\phi}\cdot \diff \bm{a}\right). 
\ee
In order to perform the $3$d Abelian duality, we add the auxiliary $\mathbb{R}^{N-1}$-valued one-form field $\bm{h}$ with the term ${g^2\over 8\pi^2 L}|\bm{h}|^2$. By shifting $\bm{h}\mapsto \bm{h}+{2\pi \im L\over g^2}\star \diff \bm{a}$, we obtain 
\be
\int_{\mathbb{R}^3}\left({1\over 2g^2 L}|\diff \bm{\phi}|^2+{g^2\over 8\pi^2 L}|\bm{h}|^2 +{\im\over 2\pi} \left(\bm{h}+{ \theta\over 2\pi}\diff \bm{\phi}\right)\cdot \diff \bm{a}\right). 
\ee
The path integral over $\bm{a}$ gives 
\be
\bm{h}+{\theta\over 2\pi}\diff \bm{\phi}=\diff \bm{\sigma}, 
\ee
where the dual photon field $\bm{\sigma}$ has the periodicity $\bm{\sigma}\sim \bm{\sigma}+2\pi \bm{\mu}_i$ for $i=1,\ldots, N-1$. In components, we can denote it as $\bm{\sigma}=\sum_i \sigma_i \bm{\mu}_i$ with $2\pi$-periodic scalars $\sigma_i$. 

Let us check if the anomaly of the $\theta$-angle periodicity can be reproduced in the dual picture. We decompose the four-dimensional $\mathbb{Z}_N$ two-form gauge fields as 
\be
B^{(2)}_{4\mathrm{d}}= A_{3\mathrm{d}}\wedge {\diff x^4\over L}+B_{3\mathrm{d}}. 
\ee
For notational simplicity, we omit the subscript $3$d in the following of this subsection. $A$ is the one-form gauge field for $\mathbb{Z}_{N}^{[0]}$ that acts on $\bm{\phi}$. In components, $\mathbb{Z}_{N}^{[0]}$ is generated by $\phi_n\mapsto \phi_n+{2\pi\over N}n$, and thus the covariant derivative takes the form 
\be
\sum_{n=1}^{N-1} (\diff \phi_n - n A) \bm{\alpha}_n=\diff \bm{\phi}-N A \bm{\mu}_{N-1}. 
\ee
To obtain the right hand side, we have used $\sum_{m=1}^{N-1}m \bm{\alpha}_m=N\bm{\mu}_{N-1}=(1,\ldots, 1,1-N)$. 
Similarly, we obtain 
\be
\sum_{n=1}^{N-1} (\diff a_n -n B)\bm{\alpha}_n=\diff \bm{a}-N B\bm{\mu}_{N-1}. 
\ee
The $3$d Abelian duality with these background gauge fields give 
\bea
S_{\mathrm{eff}}&=&\int \left({1\over 2g^2 L}\Bigl|\diff \bm{\phi}-NA\bm{\mu}_{N-1}\Bigr|^2+{g^2\over 8\pi^2 L}\Bigl|\diff \bm{\sigma}-{\theta\over 2\pi}(\diff \bm{\phi}-NA\bm{\mu}_{N-1})\Bigr|^2\right)\nonumber\\
&&-\int {\im N\over 2\pi}\bm{\mu}_{N-1}\cdot \diff \bm{\sigma}\wedge B. 
\eea
We can readily reproduce the anomaly for the $\theta$-angle periodicity of the pure Yang-Mills theory. Under the shift $\theta\mapsto \theta+2\pi$, we have to perform 
\bea
\diff \sigma_n &\mapsto& \diff \sigma_n +\bm{\alpha}_n\cdot (\diff \bm{\phi}-NA\bm{\mu}_{N-1})\nonumber\\
&=&\diff \sigma_n +\sum_{m=1}^{N-1} (\bm{\alpha}_n\cdot \bm{\alpha}_m)\diff \phi_m -\delta_{n, N-1} N A, 
\eea
in order to make the kinetic term invariant. Since $A$ is the $\mathbb{Z}_N$ gauge field, this is well defined as the transformation of the $2\pi$ periodic scalar fields $\sigma_n$. 
Substituting this transformation, we find that 
\be
S_{\mathrm{eff}}\mapsto S_{\mathrm{eff}}+{\im N\over 2\pi}\int A\wedge B,\quad (\bmod\,\, 2\pi) 
\ee
and this is nothing but the generalized anomaly between $(\mathbb{Z}_{N}^{[0]})_{3\mathrm{d}}\times (\mathbb{Z}_{N}^{[1]})_{3\mathrm{d}}$ and the $2\pi$-periodicity of $\theta$. 

Below, let us take into account the effect of the double-trace term, and we set $\rme^{\im\bm{\phi}}=\rme^{\im \bm{\phi}_\star}=(1,\omega,\ldots, \omega^{N-1})$ as the classical vacuum configuration. This process requires the gauge fixing of the remnant discrete gauge invariance by the Weyl group $W_{\mathfrak{su}(N)}$. 
The Abelianization of the gauge invariance makes the system weakly coupled, so the path integral is dominated by the saddle-point configuration in each sector. 
Such configurations of the $3$-dimensional compact Abelian gauge theory are characterized by the topological and magnetic charges, and they are called as monopole-instantons. 

There are $N$ types of the fundamental monopoles that are relevant to the analysis of YM with the double-trace deformation. $(N-1)$ of them are basically the 't~Hooft-Polyakov monopole~\cite{tHooft:1974kcl, Polyakov:1974ek}, and their effect can be written as the monopole vertex, 
\be
M_i(x)=\rme^{-S_I/N} \rme^{\im \bm{\alpha}_i\cdot \bm{\sigma}(x)} \rme^{\im \theta/N}\quad (i=1,\ldots, N-1). 
\ee
Here, $S_I={8\pi^2/g^2}$ is the instanton action, and importantly $S_I/N=8\pi^2/(g^2 N)$ remains finite in the 't~Hooft large-$N$ limit. 
The last one~\cite{Lee:1997vp, Lee:1998bb, Kraan:1998kp,Kraan:1998pm,Kraan:1998sn} corresponds to the Affine simple root, $\bm{\alpha}_N=-\bm{\alpha}_1-\ldots-\bm{\alpha}_{N-1}$, and the monopole vertex itself takes the very same form, 
\be
M_{N}(x)=\rme^{-S_I/N} \rme^{\im \bm{\alpha}_{N}\cdot \bm{\sigma}(x)}\rme^{\im \theta/N}. 
\ee
All these monopole-instantons have the topological charge $1/N$, but they are distinguished by the magnetic charge (see also \cite{Davies:2000nw}). 
Because the theory is in the weak-coupling regime, the path integral can be well approximated by the dilute gas approximation of these $N$ types of monopole-instantons.  
In order for such ensemble to make sense as a configuration on the compact $3$-dimensional space $M_3$, the total magnetic charge of that ensemble must be equal to zero. 
Importantly, the zero-mode integration of $\bm{\sigma}$ imposes this magnetic neutrality condition, so we do not need to put extra constraint to achieve it.  The magnetic neutrality also guarantees the quantization of the topological charge $\nu\in \mathbb{Z}$.  

Applying the dilute gas approximation, we obtain the monopole-instanton induced potential as  
\bea
V&=&-\rme^{-S_I/N}\sum_{n=1}^{N} \cos\left(\bm{\alpha}_n\cdot \bm{\sigma}+{\theta+\chi\over N}\right). 
\eea
Here, in order to take into account the effect of $\chi$ field, we replace $\theta$ by $\theta+\chi$. 
Solving the saddle point equation, we get the $N$ distinct saddle points, 
\be
\bm{\sigma}_k={2\pi\over N}k \bm{\rho} \equiv  {2\pi\over N}k \sum_{i=1}^{N-1}\bm{\mu}_i, 
\ee
with $k=0,1,\ldots ,N-1$ and $ \bm{\rho} $ is  the Weyl vector. 
Then, we obtain the $N$-branch structure as 
\be
\mathcal{E}_k=-N \rme^{-S_I/N} \cos\left({\theta+2\pi k+\chi\over N}\right). 
\ee

In order to obtain all the vacua of generalized YM theory, we must perform the path integral over $c^{(3)}$ and $\chi$. Path integral over $c^{(3)}$ gives 
\be
\chi={2\pi\over p}\ell \quad (\ell=0,1,\ldots, p-1), 
\ee
and we correctly obtain the $Np$-branch structure with the energy,
\be
\mathcal{E}_{\ell, k}=-N \rme^{-S_I/N} \cos\left({\theta\over N}+{2\pi (\ell + pk)\over Np}\right). 
\ee
As we have seen in the large-$N$ discussion, the label for (quasi) ground-state energies, $(\ell, k)$, cannot be simply regarded as an element of $\mathbb{Z}_p\times \mathbb{Z}_N$, but it should have an extension, 
\be
(\ell+p,k)\sim (\ell,k+1). 
\ee 
All the qualitative behaviors expected in the large-$N$ analysis are nicely obtained in the explicit evaluation of the path integral on $\mathbb{R}^3\times S^1$ with $NL\Lambda\ll 1$. 
Indeed, it has been proposed that this semiclassically computable regime, $NL\Lambda \ll 1$, is adiabatically connected to the large volume limit, $L\to \infty$, without any phase transitions~\cite{Unsal:2007jx,Unsal:2007vu,Unsal:2008ch, Poppitz:2011wy, Poppitz:2012sw,Poppitz:2012nz, Argyres:2012vv,Argyres:2012ka, Anber:2011gn, Anber:2015wha} for suitable matter contents or after double-trace  deformation.\footnote{ Adaiabatic continuity can be viewed as a stronger version of persistent order. Persistent order allows phase transitions that are admitted by mixed anomaly matching. Adiabatic continuity is the realization of a single phase at any compactification size $L$, from weak to strong coupling.}
This proposal is based on the observation that the large-$N$ volume independence works so long as the system is in the confined phase~\cite{Eguchi:1982nm, GonzalezArroyo:1982hz,GonzalezArroyo:1982ub, GonzalezArroyo:2010ss}, and the story is that the large-$N$ volume independence is true for $L\gtrsim 1/\Lambda$ and it has a smooth transition to the above semiclasically confined regime, $L\ll 1/ N\Lambda$. 
Even though this adiabatic continuity to the semiclassical regime is still an open issue, let us point out that a recent numerical lattice study~\cite{Bonati:2018rfg} of the topological susceptibility on $\mathbb{R}^3\times S^1$ with the double-trace deformation shows the evidence of volume independence, so it seems that the first half of the story is being verified. 

\section{Vacuum structure of generalized super Yang-Mills theory}\label{sec:vaccum_mSYM}

Here, we introduce  one adjoint Weyl fermion: 
\be
\overline{\lambda} \overline{\sigma}^\mu \left(\p_\mu \lambda+\im [a_\mu,\lambda]\right). 
\ee
For $p=1$, this is the $\mathcal{N}=1$ super Yang-Mills theory. 
This theory has $\mathbb{Z}_{2N}$ discrete chiral symmetry, and there are $N$ vacua associated with the chiral symmetry breaking, $\mathbb{Z}_{2N}\to \mathbb{Z}_2$~\cite{Witten:1982df}.  
In the following, we consider the case $p>1$, by restricting the instanton sum.

\subsection{$\mathbb{Z}_{2Np}$ chiral symmetry and mixed 't Hooft anomaly}

Let us identify the $0$-form symmetry, which we will call chiral symmetry. Under the rotation $\lambda\mapsto \rme^{\im \alpha}\lambda$, often denoted as $U(1)_R$, the fermion measure gives the contribution 
\be
\im \alpha{2N \over 8\pi^2}\int \tr (F\wedge F). 
\ee
We can combine this with another $U(1)$ transformation $\chi\mapsto \chi+\beta$, and then we get the additional term as 
\bea
&&\im {2N \alpha\over 8\pi^2}\int \tr (F\wedge F)+\im \beta \int\left({1\over 8\pi^2}\tr (F\wedge F)-{p\over 2\pi} \diff c^{(3)}\right)\nonumber\\
&=& \im {2N \alpha + \beta \over 8\pi^2}\int \tr (F\wedge F) - \im {p \beta \over 2\pi}\int \diff c^{(3)}. 
\eea
In order for this to be a symmetry, we require that 
\be
2N \alpha + \beta \in 2\pi \mathbb{Z},\; p \beta\in 2\pi \mathbb{Z}. 
\ee
This is solved as 
\be
\alpha = {2\pi \over 2Np} n ,\, \beta = -{2\pi\over p}n, 
\ee
with $n\sim n +2 Np$. This shows that we have $\mathbb{Z}_{2N p}$ discrete chiral symmetry, generated by 
\be
\lambda\mapsto \rme^{2\pi \im\over 2N p}\lambda, \; \rme^{\im\chi} \mapsto \rme^{-{2\pi\over p}\im}\rme^{\im \chi}. 
\ee
Therefore, the chiral symmetry is enlarged by $\mathbb{Z}_p$. 
The reason for this enhancement is that we can combine the discrete symmetry by a combination of two quantum-mechanically broken $U(1)$ symmetries, $U(1)_R$ and the Peccei-Quinn $U(1)$ symmetry~\cite{Peccei:1977hh,Peccei:1977ur}. 
As in the $p=1$ case, the subgroup $\mathbb{Z}_2\subset \mathbb{Z}_{2Np}$ is the fermion parity, and it cannot be broken in the Lorentz-invariant vacuum. 

For the minimal generator of $\mathbb{Z}_{2Np}$, 
\be
\alpha={2\pi\over 2Np}, \; \beta=-{2\pi\over p}, 
\ee
the theta angles $\theta_{\mathrm{YM}}$ and $\hat{\theta}$ are shifted as 
\be
\theta_{\mathrm{YM}}\mapsto \theta_{\mathrm{YM}}+2N\alpha+\beta=\theta_{\mathrm{YM}},\; 
\hat{\theta}\mapsto \hat{\theta}-p\beta=\hat{\theta}+2\pi. 
\ee
As we have discussed in Sec.~\ref{sec:mixed_anomaly_theta}, the $\theta$-angle periodicity has a mixed anomaly with $\mathbb{Z}_{N}^{[1]}\times \mathbb{Z}_{p}^{[3]}$. 
In the case of SYM, this leads to the mixed 't~Hooft anomaly with the discrete chiral symmetry. By introducing the background $U(1)$ higher-form gauge fields with the constraint
\be
N B^{(2)}=\diff B^{(1)},\; p D^{(4)}=\diff D^{(3)}+{N\over 4\pi} B^{(2)}\wedge B^{(2)}, 
\ee
we find that the discrete chiral transformation gives the phase rotation of the partition function as 
\be
Z\mapsto  \exp \left(-\im \int D^{(4)}\right)Z. 
\ee
This shows that the system has $\mathbb{Z}_{Np}$ 't Hooft anomaly. 
We will see in the following that the anomaly is matched by the discrete chiral symmetry breaking $\mathbb{Z}_{2Np}\to \mathbb{Z}_2$.

\begin{figure}[t]
\vspace{-2cm}
\begin{center}
\hspace{0cm} \includegraphics[width = 0.8 \textwidth]{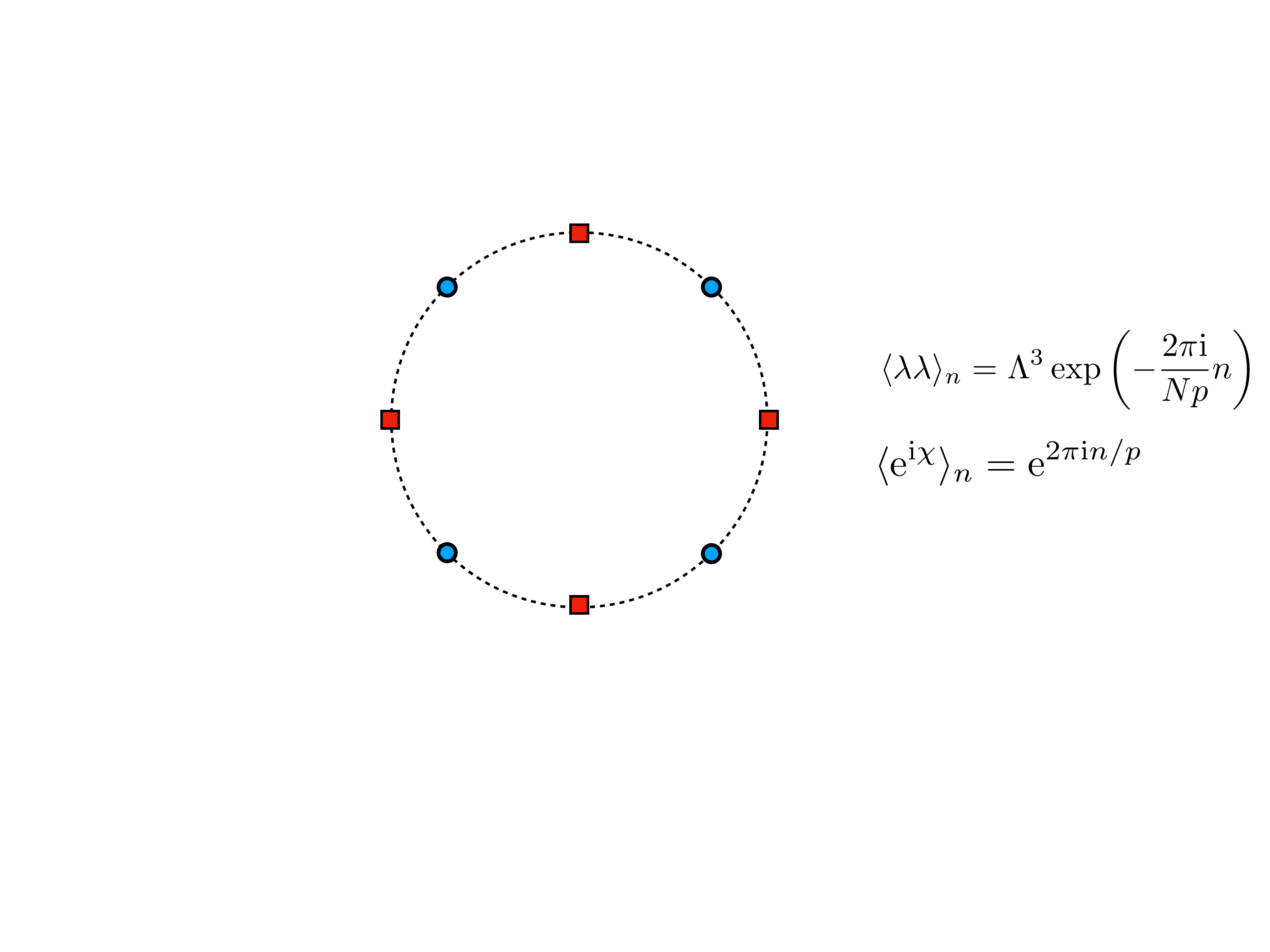}
\vspace{-3.0cm}
\caption{The vacuum structure of $SU(N)$  SYM   theory where instanton sum is restricted to multiples of charge-$p$ 
($N=4, p=2$ in the figure). 
 The theory has $Np$ vacua. These vacua split naturally  to two sets, denoted by the blue circles and the red squares. The vacua for which $n_1 -n_2 =0$ mod $p$ are relative superselection sectors. There are  dynamical domain walls in between. The vacua for which $n_1 -n_2 \neq 0$ mod $p$ are different  {\it universes}. 
There exists no dynamical domain walls that can connect them, and no tunneling in between exists  even when   the theory is compactified.  Only external probes charged under  $\mathbb{Z}_{p}^{[3]}$ can connect them. When mass deformation is added, this structure extrapolates to $Np$ branches of generalized YM theory. 
 }
\label{fig:vacua}
\end{center}
\end{figure}

\subsection{Symmetry breaking, superselection sectors, and universes}\label{sec:superselection_universes}

In this section, we discuss the consequence of the 't~Hooft anomaly matching. We can set $\theta_{\mathrm{YM}}=\hat{\theta}=0$. 
Since the path integral over $c^{(3)}$ can be done exactly, we find that 
\be
\chi\in{2\pi\over p}\mathbb{Z}. 
\ee
For each fixed $\chi$, the path integral over $a$ and $\lambda$ is identical with the usual SYM with $\theta_{\mathrm{YM}}=\chi$, so we obtain the $N$ vacua distinguished by the gluino condensate, 
\be
\langle \lambda \lambda\rangle=\Lambda^3 \exp\left(-\im {\chi\over N}+{2\pi\im\over N}k\right). 
\ee
Substituting $\chi={2\pi\over p} \ell$, we find that the theory has $Np$ vacua characterized by the condensates,
\be
\langle \lambda \lambda\rangle_n=\Lambda^3 \exp\left(-{2\pi\im \over Np}n\right),\; \langle \rme^{\im \chi}\rangle_n=\rme^{2\pi\im n/p}, 
\label{eq:vacua_mSYM}
\ee
for $n=0,1,\ldots, Np-1$. Here, we combine the label $(\ell,k)$ as a single integer, $n=\ell+pk$, because of the identification $(\ell+p,k)\sim (\ell,k+1)$. 
This is shown in Fig.~\ref{fig:vacua}.

Since $\mathbb{Z}_{2Np}$ discrete chiral symmetry is spontaneously broken to $\mathbb{Z}_2$, we are tempted to conclude that there are $Np$ superselection sectors.  
However, the existence of the $\mathbb{Z}_p$ three-form symmetry tells us a more strong rule, and we  call them as different universes.

To explain it, let us start with the explanation of superselection rule. Let $\mathcal{H}$ be the Hilbert space of a given QFT, then we say that 
\begin{itemize}
\item $|\Phi_1 \rangle, |\Phi_2\rangle\in\mathcal{H}$ are distinguished by superselection rule. \\
$\Leftrightarrow$ For any local observables $O$, $\langle \Phi_1| O |\Phi_2\rangle=0$.
\end{itemize}
The definition of superselection rule crucially depends on the choice of allowed observables. 
In the context of local QFT, the space of observables is usually taken to be the set of local operators, and we also adopt this convention here. 
We therefore have the superselection sectors by spontaneous symmetry breaking, because the transition from one symmetry-broken vacuum to another is an extensive operation while the local operator cannot create such extensive energy. 
In order to go to another vacuum, we need a nonlocal operation such as heating up the whole magnet above the critical temperature and cooling it down. 

In the case of QFT, the superselection rule usually appears only in the infinite volume limit. 
In the finite volume, the symmetric state always has a lower energy than asymmetric states by an exponentially small energy\footnote{In some QFTs, the domain wall may be associated with some fermionic zero modes under a suitable boundary condition, and then it does not lead the energy splitting at the finite volume. Even in such cases, we can prepare a certain local point-like operator to find the would-be exponentially small energy splitting, and we can notice the existence of dynamical domain wall. As we will explain, the distinction as universes is stronger than such situations.}. The physical interpretation of this is the existence of domain wall as a dynamical excitation connecting different symmetry-broken vacua. 

In the case of generalized YM, the situation is different because of the topological coupling, 
\be
{\im p\over 2\pi}\int \chi\wedge \diff c^{(3)}. 
\label{eq:EFT_chi}
\ee 
The path integral over $c^{(3)}$ exactly requires that $\chi$ must be a constant modulo $2\pi$. 
Therefore, there does not exist any configurations $\chi(t)$ connecting different vacua, such as 
\be
\chi(-\infty)=0, \, \chi(+\infty)={2\pi\over p}. 
\ee
The path integral with this boundary condition is exactly zero even in the finite volume. This conclusion does not change after inserting any local point-like operators. 
Therefore, at any finite volume,  the off-diagonal matrix elements for two states, $\chi=0$ and $\chi={2\pi\over p}$, of any local operator vanishes, $\langle \chi=0| O(x)| \chi=2\pi/p\rangle=0 $. 
That is, the superselection rule persists to the finite volume if two states have different values of the scalar field $\chi$. 
This superselection rule is clearly stronger than the ordinary one, and  to emphasize the distinction, we  call them  different universes. 

We call two states  $|\Phi_1\rangle, |\Phi_2\rangle\in \mathcal{H}$ as different universes if 
\begin{itemize}
\item Even in the finite volume, the superselection rule distinguishes them:\\ $\langle \Phi_1| O(x) |\Phi_2\rangle =0$ for any local observable $O(x)$ and any finite volume. 
\item There is no dynamical domain wall connecting those two states, $\Phi_1$ and $\Phi_2$. 
\end{itemize}
We believe that these two statements are equivalent for local QFT. 

In the case of generalized SYM, there is a good symmetry reason why there are different universes. 
Because of the mixed anomaly between the chiral symmetry $\mathbb{Z}_{2Np}$ and $\mathbb{Z}_p$ $3$-form symmetry, the discrete chiral transformation by one unit generates the $\mathbb{Z}_p$ phase. 
Therefore, the wall configuration connecting two vacua, $n=0$ and $n=1$, has to be charged under the $\mathbb{Z}_p$ three-form symmetry. 
However, the existence of three-form symmetry means that there is no dynamical wall excitations charged under it. 
Therefore, the mixed anomaly between $0$-form and $3$-form symmetries ensures that the symmetry broken vacua for the $0$-form symmetry are distinguished as different universes and superselection sectors. 
This fact can be generalized to $d$-dimensional QFT if it has a mixed anomaly between $0$-form and $(d-1)$-form symmetries. 
Indeed, in the case of $d=2$,  charge-$p$ models discussed in Refs.~\cite{Anber:2018jdf,Anber:2018xek, Armoni:2018bga, Misumi:2019dwq, Cherman:2019hbq} can be interpreted 
as  realizations of different universes by having a $0$-form and $1$-form mixed anomaly.

The two vacua are different universes if $\chi$ takes different values in $\bmod\,\, 2\pi$.  More explicitly, if we consider two vacua $n=n_1$ and $n_2$ in (\ref{eq:vacua_mSYM}), then they are different universes if and only if $n_1\not=n_2$ mod $p$. 
If they are different in mod $p$, the wall configuration is possible only after we insert the external probe,
\be
\exp\left(\im (n_1-n_2)\int c^{(3)}\right). 
\ee
The insertion of this probe shifts $\chi \to \chi+{2\pi\over p}(n_1-n_2)$ $\bmod\, 2\pi$, and these configurations are related by spontaneously broken chiral symmetry, so the wall configurations connecting different universes are obtained by insertion of this probe operator. As a consequence, unlike generalized YM, the $\mathbb{Z}_p^{[3]}$ symmetry is spontaneously broken in generalized SYM, and one can confirm with the criterion (\ref{eq:order_3form}). 
When $n_1=n_2+p k$, there is a dynamical domain wall connecting those vacua, and it is consistent with the fact that the above operator becomes neutral under $\mathbb{Z}_{p}^{[3]}$. 
If $k$ is not a multiple of $N$, then the mixed anomaly between chiral and one-form symmetries shows that the effective field theory on the domain wall has an 't~Hooft anomaly for the $\mathbb{Z}_N$ one-form symmetry. 
For the study of the physics on the wall, see, e.g., Refs.~\cite{Gaiotto:2017yup, Anber:2015kea, Sulejmanpasic:2016uwq, Komargodski:2017smk, Anber:2018jdf,Anber:2018xek, Nishimura:2019umw, Cox:2019aji}. 

It is important to emphasize that the absence of dynamical domain wall is completely characterized by the mixed anomaly between a $0$-form symmetry and a $(d-1)$-form symmetry in $d$-dim QFTs. 
In the case of generalized YM theory, it comes out of TQFT (\ref{eq:EFT_chi}). Therefore, in the $ \mathbb{Z}_{2Np} \rightarrow  \mathbb{Z}_2 $ chiral symmetry breaking pattern, the  breaking of the $\mathbb{Z}_{p}\simeq \mathbb{Z}_{2Np}/\mathbb{Z}_{2N}$ part of chiral symmetry  can be viewed as a UV (or all-scale) phenomenon  due to its TQFT nature,   leading to different  universes. 
The breaking of $\mathbb{Z}_{N}$ is a dynamical IR phenomenon, leading to superselection sectors within each universe, and these two chiral-symmetry breakings have different physical origins. 

An interesting implication of this is that 
the  distinction between superselection sectors and universes can be emergent in the low-energy effective description if we introduce a finite UV cutoff. 
For example, the action (\ref{eq:EFT_chi}) can be regarded as the low-energy effective description of 
\be
\int {1\over 2}\left(v^2|\diff \chi|^2+ \lambda(1- \cos (p\chi))\right). 
\ee
Here, $v$ and $\lambda$ are dimensionful parameters, with mass dimensions $[v]=1$ and $[\lambda]=4$, and this theory is not renormalizable because of the cosine-type interaction. We assume that this is already some low-energy effective description of a UV complete theory, but, for our purpose, such details are unimportant. 
This model has the ordinary $\mathbb{Z}_{p}$ symmetry as $\chi \mapsto \chi+2\pi/p$, but there is no $3$-form symmetry. 
In this case, the three-form symmetry of (\ref{eq:EFT_chi}) is an emergent symmetry in the low-energy effective description. 
The fact that the wall is charged under the emergent $\mathbb{Z}_p$ $3$-form symmetry tells us that the dynamical domain wall may be possible in the UV description, but the energy density of such wall is controlled by the typical energy scale of that UV theory. 
For example, in the specific UV theory given above, the domain-wall energy density is characterized by $v\sqrt{\lambda}/p$, and it goes to infinity as sending the parameters of UV theory to infinity, $v,\lambda\to \infty$. 
This has the sharp contrast with the fact that the energy density of the dynamical domain walls (i.e., when $n_1-n_2=0 \bmod p$) is controlled by the strong scale of ${\cal N}=1$ SYM.

\section{Generalized QCD with fundamental fermion}\label{sec:QCD_fund}

Let us briefly discuss the modified instanton sum in QCD with one-flavor fundamental quark, 
\be
\overline{\psi}\gamma^\mu (\p_\mu + \im a_\mu)\psi. 
\ee
At the classical level, the theory has the $U(1)_A$ axial symmetry. In the ordinary $1$-flavor QCD, the instanton contribution breaks this axial symmetry completely, and thus the additive mass renormalization exists. 
However, the massless point is still a well-defined point at least within the continuum formulation, since the additive renormalization by instanton contribution is softer than the mass effect in the high momentum limit~\cite{Gaiotto:2017tne}. 

The special feature of the massless point becomes very evident when we consider the modified instanton sum. 
In order to see it, let us consider the axial transformation $\psi\mapsto \rme^{\im \alpha \gamma_5}\psi$ and $\chi\mapsto \chi+\beta$, then the total change of the action is given by  
\be
2\im \alpha {1\over 8\pi^2}\int \tr (F\wedge F) + \im \beta \int \left({1\over 8\pi^2}\tr (F\wedge F) + {p\over 2\pi} \diff c^{(3)}\right). 
\ee
Therefore, this transformation is a symmetry if
\be
2\alpha + \beta \in  2\pi \mathbb{Z},\; p \beta \in 2\pi \mathbb{Z}. 
\ee
This shows that modified instanton sum in massless $1$-flavor QCD provides $\mathbb{Z}_{2p}$ discrete chiral symmetry, and it is generated by 
\be
\psi \mapsto \exp\left({2\pi\over 2p}\im \gamma_5\right)\psi,\quad \rme^{\im \chi}\mapsto \exp\left(-{2\pi\over p}\im\right) \rme^{\im\chi}. 
\ee
The subgroup $\mathbb{Z}_2\subset \mathbb{Z}_{2p}$ is the fermion parity, so it cannot be spontaneously broken in Lorentz-invariant vacuum. 
We will see that this discrete chiral symmetry is spontaneously broken as 
\be
\mathbb{Z}_{2p} \xrightarrow{\mathrm{SSB}} \mathbb{Z}_2, 
\ee
and we have $p$ distinct vacua. Moreover, they are different universes. 

In order to see it, we discuss the mixed anomaly between the discrete chiral symmetry and the higher-form symmetry. 
Since the pair creation and annihilation of fundamental quarks break the string of Wilson lines, this theory does not have $\mathbb{Z}_{N}^{[1]}$. Therefore, we must turn off $B^{(2)}$ and $B^{(1)}$ in the analysis of Sec.~\ref{sec:mixed_anomaly_theta}, and thus the background gauge fields $D^{(4)}$ satisfies 
\be
p D^{(4)}=\diff D^{(3)}. 
\ee
This is the background gauge field for $\mathbb{Z}_p^{[3]}$ acting on $V^{(3)}=\exp(\im \int c^{(3)})$. 
By performing the $\mathbb{Z}_{2p}$ discrete chiral transformation under the existence of $D^{(4)}$, the partition function transforms as 
\be
Z\mapsto \exp\left(\im \int D^{(4)}\right) Z. 
\ee
Therefore, there is the $\mathbb{Z}_p$ mixed 't~Hooft anomaly for $\mathbb{Z}_{2p}^{[0]}\times \mathbb{Z}_{p}^{[3]}$. 
This anomaly is matched by the spontaneous breaking of discrete chiral symmetry. 
Also, this mixed anomaly tells us that superselection rule between different symmetry-broken vacua holds true for any finite volumes, and no dynamical domain wall can interpolate between them.

  It should be noted that   this construction itself does not provide a framework in which $\mathbb Z_{2p}$ symmetry protects fermion mass operator from being generated, despite the fact that it forbids the mass operator, $\overline{\psi}  \psi$.  How is this possible? 
  For example, small instanton amplitudes of the form 
$e^{-S_I} \overline{\psi} \rm P_{\rm R} \psi$  is  forbidden due to $\mathbb{Z}_{2p} $ but  $e^{-S_I} \overline{\psi} \rm P_{\rm R} \psi \rme^{\im \chi} $ is allowed.   As emphasized earlier, 
the vev  $ \langle \rme^{\im \chi} \rangle_n \neq 0 $  is  an all scale property, valid at both UV and IR due to topological coupling \eqref{eq:EFT_chi}. Because of this reason, formally, the chiral symmetry breaking scale via this operator may be viewed as UV energy scale, and a soft  mass term is generated due to instantons in any one of the $p$ universes.  
However, it is possible that our generalized QCD can emerge as an effective description of another QFT, with emergent $3$-form symmetry at low energies. 
Even in such cases, the domain-wall excitations or the false-vacuum decays are exponentially suppressed because of the emergent $3$-form symmetry, so our consideration here may lead to  interesting physical consequences. 

\section{Connection between modified instanton sums in $4$d and $2$d}\label{sec:4d_2d}

So far, we have discussed the effect of modified instanton sum in the context of $4$d Yang-Mills theory and QCD, 
revealing the implications of the proposal in \cite{Seiberg:2010qd}. There are works  
 in the context of $2$d charge-$p$  Schwinger   and   sigma models which can be interpreted as modified instanton sums: 
 ~\cite{Pantev:2005zs,Pantev:2005rh, Pantev:2005wj, Sharpe:2014tca,  Anber:2018jdf,Anber:2018xek, Armoni:2018bga, Misumi:2019dwq, Sharpe:2019ddn}  and indeed, very similar vacuum structures can be found there. 
Especially in \cite{Sharpe:2014tca}, using the terminology of this paper, it has been observed that the path integral of charge-$p$ $\mathbb{C}P^{N-1}$ model has the decomposition into $p$ universes, and the summation over those universes give the projection operator in $2$ dimensions.  

The purpose of this section is to uncover the possible connection between these  two class of theories (in 4d vs. 2d)  in view of anomaly, following the idea of Refs.~\cite{Shimizu:2017asf, Tanizaki:2017mtm,Tanizaki:2017qhf, Yamazaki:2017dra, Dunne:2018hog}. 
We note that, applying the discussion of Ref.~\cite{Yamazaki:2017ulc}, we can obtain more explicit connection between generalized YM theory on $T^3\times \mathbb{R}$ with 't~Hooft flux and the charge-$p$ $\mathbb{C}P^{N-1}$ model on $\mathbb{R}\times S^1$ with $\mathbb{Z}_N$ twisted boundary condition. 
The $p=1$ version of the twisted $\mathbb{C}P^{N-1}$ model was proposed to show the volume independence in Refs.~\cite{Dunne:2012ae, Dunne:2012zk}, and it (and its cousins) have been discussed in details in Refs.~\cite{Sulejmanpasic:2016llc, Misumi:2014jua, Misumi:2014bsa,Fujimori:2017osz,Fujimori:2018kqp, Cherman:2013yfa,Cherman:2014ofa, Evslin:2018yfm, Hongo:2018rpy}. 

We consider the compactification on two torus $T^2=(S^1)_{L_1}\times (S^1)_{L_2}$ of $4$d generalized YM theory, and discuss the anomaly of the effective $2$-dimensional field theory. We denote our four-dimensional spacetime as $M_4=(S^1)_{L_1}\times (S^1)_{L_2}\times M_2$, and the size of $M_2$ is much larger than $(S^1)_{L_1}\times (S^1)_{L_2}$. 
In order to obtain the anomaly of effective $2$d theory, it is sufficient to identify the contents of background gauge fields on $M_2$ induced from $M_4$. 
First, we decompose $B^{(2)}$ as 
\be
B^{(2)}=A_1 \wedge {\diff x_1\over L_1}+A_2\wedge {\diff x_2\over L_2} + B,
\ee
where $A_{1}$ and $A_2$ are $\mathbb{Z}_N$ one-form gauge field and $B$ is $\mathbb{Z}_N$ two-form gauge field on $M_2$. 
Therefore, 
\be
{N\over 4\pi}\int_{T^2} B^{(2)}\wedge B^{(2)}=-{N\over 2\pi}A_1\wedge A_2,  
\ee
and this provides the $\mathbb{Z}_N$ two-form gauge field. 

Next, we consider the compactification of $D^{(4)}$ and $D^{(3)}$ as 
\be
D^{(4)}= D\wedge {\diff x_1\wedge \diff x_2\over L_1\, L_2},\; D^{(3)}=C\wedge {\diff x^1\wedge \diff x^2\over L_1 \, L_2}. 
\ee
As a compactification of the constraint equation (\ref{eq:4group}), we obtain 
\be
p D =\diff C-{N\over 2\pi} A_1\wedge A_2. 
\ee
As a summary, by considering $T^2$ compactification, the higher-form symmetry is converted as 
\be
\Bigl(\mathbb{Z}_N^{[1]}\times \mathbb{Z}_p^{[3]}\Bigr)_{4\mathrm{d}}\, \Rightarrow\, \Bigl(\mathbb{Z}_{N}^{[0]}\Bigr)_1\times \Bigl(\mathbb{Z}_{N}^{[0]}\Bigr)_2\times \mathbb{Z}_{N}^{[1]}\times \mathbb{Z}_{p}^{[1]}.  
\ee
$A_1$ and $A_2$ are the background gauge fields for $\Bigl(\mathbb{Z}_{N}^{[0]}\Bigr)_{1,2}$, $D$ is the two-form gauge field for $\mathbb{Z}_{p}^{[1]}$, and the flux of $D$ is affected by the $0$-form symmetry. 
This shows that, under the shift of $\theta_{\mathrm{YM}}\to \theta_{\mathrm{YM}}+2\pi/p$, we obtain the anomalous phase as 
\be
Z_{M_2}[A_1,A_2,D]\mapsto \exp\left(-\im \int D\right) Z_{M_2}[A_1, A_2, D]. 
\ee
This is indeed the slight generalization of the anomaly of $\mathbb{C}P^{N-1}$ model computed in Ref.~\cite{Komargodski:2017dmc} for the charge-$p$ model and by restricting the $PSU(N)$ symmetry to $\mathbb{Z}_N\times \mathbb{Z}_N$~\cite{Tanizaki:2018xto}. 
This explains why $2$d charge-$p$ $\mathbb{C}P^{N-1}$ model studied in Refs.~\cite{Pantev:2005zs,Pantev:2005rh, Pantev:2005wj, Seiberg:2010qd} shows  very similar vacuum structure with that of generalized YM theory. 

\section{Conclusions and Outlooks}
The main results of this work can be summarized as follows: 

\begin{itemize} 
\item{Modifying instanton sums into the integer multiples of topological charge $p$ can be established within local QFTs. This modification is realized by coupling a  TQFT to QFT. 
In the case of $4$d $SU(N)$ YM theory, this process naturally creates a $3$-form symmetry.   }

\item{ In general, $(d-1)$-form symmetries in $d$ spacetime dimensions  cannot be spontaneously broken in the absence of  mixed anomalies.    If broken due to a mixed anomaly,  we will always obtain  universes instead of superselection sectors.}

\item{In generalized Yang-Mills theory, we showed mixed anomaly between $\mathbb{Z}_{N}^{[1]}\ttimes \mathbb{Z}_{p}^{[3]}$ symmetry and the $\theta$-angle periodicity. Gauging this symmetry correctly necessitates an intriguing  4-group structure. The vacuum has $Np$ branches. }

\item{Generalized SYM has $Np$ isolated vacua and 1-flavor QCD has $p$ vacua. The vacua separated by $n_1 - n_2= 0$ mod $p$ units are interpreted  as superselection sectors,  while the ones for which  $n_1 - n_2 \neq 0$ mod $p$ are  referred to as universes. There are no dynamical domain walls connecting them.  But the wall configuration can be generated in the presence of an external probe charged under  $\mathbb{Z}_{p}^{[3]}$.}

\item{The distinction as universes appears if the broken $0$-form symmetry has a mixed anomaly with a $(d-1)$-form symmetry in $d$-dim QFT. If the would-be domain wall connecting distinct vacua is charged under the $(d-1)$-form symmetry, such configuration cannot arise as a dynamical excitation. }

\item{
We proved  compatibility of abelian duality with mixed anomaly, in the context of  center-stabilized YM  theory  on $\mathbb R^3 \times S^1$. Both original and dual formulation produce the same  mixed anomaly polynomials.\footnote{From many examples of  semi-classics (within its regime of applicability) on $\mathbb R^3 \times S^1$,  and mixed anomalies, it is already  known that the two never contradict each other.   Our construction goes one step beyond this point  and derives the anomaly polynomial of the original theory  from the dual  semi-classical formulation.}
For the generalized YM, the dilute gas of monopole-instantons in the semi-classical regime produces the correct $Np$ branched structure.}

\item{Generalized YM  theory, upon compactification on $T^2 \times { M}_2$, reduce to charge-$p$   ${\mathbb CP}^{N-1}$ model. This uncovers the close connection between modified instanton sums in $4$d and $2$d field theories.  }
\end{itemize}

We still have many open issues related to this study. Let us briefly summarize our outlook for possible extensions: 
\begin{itemize}
\item We should look for physical applications of the notion of universes. We describe an outlook on this point later in more detail. 
\item We have studied the modified instanton sum for $SU(N)$ gauge theories. Generalization to the non-simply connected gauge groups needs to be explored. 
\item We derived the anomaly of $3$d dual theory for deformed YM on $\mathbb{R}^3\times S^1$, however it is not immediately obvious how we can generalize to other compactified gauge theories on $\mathbb{R}^3\times S^1$. Especially, it is important to study how we can concretely obtain the anomaly of compactified $\mathcal{N}=1$ SYM in a consistent way with known effective (Affine Toda) superpotential based on monopoles. 
\item In $2$d, the modified instanton sum has a nice application to define the lattice $\theta$ angle~\cite{Gattringer:2018dlw, Sulejmanpasic:2019ytl}. Does it have a generalization to $4$d Yang-Mills theory? 
\end{itemize}

It would be especially interesting to apply the observations of this study into realistic physical situations. Our study shows that if one QFT couples to other hidden sectors with a $(d-1)$-form symmetry, then the total system has the decomposition into distinct sectors, called universes. 
We hope that this observation has an interesting application to the false vacuum to decay.  
Let us assume that the system is trapped by the false vacuum distinguished as the universe from the true ground state, then the false-vacuum decay is prohibited. In order for the false vacuum decay, the system has to create bubbles of the true vacuum, which is nothing but the wall configuration between the true and false vacua. Then, the boundary of the bubble must be charged under the $(d-1)$-form symmetry, and this means that it can appear only after insertion of a $(d-1)$-dimensional external probe. 
When the $(d-1)$-form symmetry is emergent, this selection rule is true only in the approximate sense. Still, it means that the system has to create very large bubbles of the true vacuum in order for the bubble to expand by the pressure difference between inside and outside of the bubble. We can expect that the life-time of the false vacuum becomes much longer than usual cases. 

\acknowledgments

The authors thank Zohar Komargodski, Erich Poppitz and Eric Sharpe for many useful discussions. 
The work of Y.~T. is supported by the JSPS Overseas Research Fellowships. 
The work of M.~\"{U} is supported by the U.S. Department of Energy, Office of Science, Office of Nuclear Physics under Award Number DE-FG02-03ER41260.

\appendix

\section{Explicit formulas of roots and weights of $\mathfrak{su}(N)$}\label{sec:su(N)_root_weight}
$\mathfrak{su}(N)$ is the Lie algebra of the $N\times N$ Hermitian, traceless matrices. Let $\bm{e}_i$ be the $i$-th $N$-dimensional unit vector, then the canonical basis of $\mathfrak{su}(N)_{\mathbb{C}}$ is given by the Cartan generator $\{H_m\}_{m=1,\ldots,N-1}$ and the root $\{E_{\bm\alpha_{ij}}\}_{1\le i\not=j\le N}$:
\be
H_m={1\over \sqrt{m(m+1)}}\left(\sum_{i=1}^m \bm e_i\otimes \bm e_i-m \bm e_{m+1}\otimes \bm e_{m+1}\right),\; E_{\bm\alpha_{ij}}=\bm{e}_i\otimes \bm{e}_j. 
\ee
It satisfies 
\be
\tr[H_i H_j]=\delta_{ij},\; \tr[E_{\bm\alpha_{ij}}E_{\bm\alpha_{k\ell}}]=\delta_{i\ell}\delta_{jk}, 
\ee
where the trace is taken for the defining representation (i.e. the faithful representation with the smallest dimension), and this makes the length squared of root as $\bm \alpha^2=2$. 
The commutation relation is 
\bea
[H_m,E_{\bm\alpha_{k\ell}}]&=&{(\bm{e}_k-\bm{e}_\ell)\cdot(\bm{e}_1+\cdots+\bm{e}_m-m\bm{e}_{m+1})\over \sqrt{m (m+1)}} E_{\bm\alpha_{k\ell}}. 
\eea
Since the vectors $(\bm{e}_1+\cdots+\bm{e}_m-m\bm{e}_{m+1})/\sqrt{m(m+1)}$ form the orthonormal basis of the $(N-1)$-dimensional subspace of $\mathbb{R}^N$, this commutation relation tells us that the root vector of $E_{\bm\alpha_{k\ell}}$ is 
\be
\bm{\alpha}_{k\ell}=(\bm{e}_k-\bm{e}_\ell),
\ee
and we call they are positive if $k<\ell$. The simple roots are 
\be
\bm{\alpha}_i=\bm{\alpha}_{i\, i+1}=(\bm{e}_i-\bm{e}_{i+1}),
\ee 
and then $\bm\alpha_{k\ell}=\bm\alpha_k+\cdots+\bm\alpha_{\ell-1}$ for $k<\ell$. In this convention, it is now evident that $|\bm\alpha_{k\ell}|^2=2$, and thus we can identify the co-root/root and co-weight/weight as $\bm\alpha_i^{\vee}=\bm\alpha_i$ and $\bm\mu_i^{\vee}=\bm\mu_i$, respectively (i.e. $\Lambda_R^{\vee}=\Lambda_R$ and $\Lambda_W^{\vee}=\Lambda_W$). We can check that the Cartan matrix is of the class $A_{N-1}$, 
\be
C_{ij}:={2(\bm\alpha_i\cdot \bm\alpha_j)\over |\bm\alpha_j|^2}=\bm\alpha_i\cdot \bm\alpha_j^{\vee}=2\delta_{i j}-\delta_{i\, j+1}-\delta_{i+1\, j}. 
\ee
The inner product between the root space and $\bm H$, such as $\bm\alpha_i\cdot \bm H$, is understood as 
\bea
\bm\alpha_i\cdot \bm H&=&\sum_{m=1}^{N-1}{\bm\alpha_i\cdot(\bm{e}_1+\cdots+\bm{e}_m-m\bm{e}_{m+1})\over \sqrt{m (m+1)}}H_m\nonumber\\
&=&-\sqrt{i-1\over i}H_{i-1}+\sqrt{i+1\over i}H_i=\bm e_i\otimes \bm e_i-\bm e_{i+1}\otimes \bm e_{i+1}, 
\eea
and then $[E_{\bm\alpha_i},E_{-\bm\alpha_i}]=\bm\alpha_i \cdot \bm H$ holds. 

To write down the fundamental weights, it is convenient to introduce the following vectors in the root space, 
\be
\bm{e}'_i=\bm{e}_i-{1\over N}\sum_{k=1}^{N}\bm{e}_k. 
\ee
By definition, $\bm{e}'_i={1\over N}\sum_j \bm\alpha_{ij}\in \sum_{k=1}^{N-1} \bm\alpha_k \mathbb{R}$, and $\bm{e}'_i\cdot \bm\alpha_j=\bm{e}_i\cdot \bm\alpha_j=\delta_{ij}-\delta_{i\, j+1}$. The fundamental weights are given by 
\be
\bm{\mu}_i=\sum_{j=1}^{i}\bm{e}'_i, 
\ee
and thus $\bm{e}_i'=\bm\mu_i-\bm\mu_{i-1}$ setting $\bm{\mu}_0=\bm{\mu}_N=0$. Therefore, the simple roots can be written as 
\be
\bm\alpha_i=\bm e'_i-\bm e'_{i+1}=2\bm{\mu}_i-\bm{\mu}_{i+1}-\bm{\mu}_{i-1}. 
\ee

Using this property, let us compute $\Lambda_W/\Lambda_R(=\Lambda_W^{\vee}/\Lambda_{R}^{\vee})$ explicitly. Let $\bm\lambda=\sum_i\lambda_i \bm\mu_i$ be a weight. In $\Lambda_W/\Lambda_R$, we regard $\bm\alpha_i\sim 0$, which can be solved as $\bm\mu_n\sim n\bm\mu_1$ and $N\bm\mu_1\sim 0$ for fundamental weights. 
As a result, 
\be
\bm\lambda=\sum_n \lambda_n\bm\mu_n\sim \sum_n (n\lambda_n) \bm\mu_1
\ee
and the coefficient is further identified by modulo $N$. Therefore, 
$\Lambda_W/\Lambda_R\simeq \mathbb{Z}_N=Z(SU(N))$ and the isomorphism is given by the mapping 
\be
(\lambda_1,\ldots,\lambda_{N-1})\mapsto \sum_n n \lambda_n\bmod N,
\ee
which is called $N$-ality.

\bibliographystyle{utphys}
\bibliography{QFT}
\end{document}